\documentclass[conference,a4paper]{APSIPA2021}
\usepackage{multirow}
\usepackage[dvipdfmx]{graphicx}
\usepackage{amsmath}
\usepackage[psamsfonts]{amssymb}
\usepackage{subfigure}
\usepackage{amsxtra}
\usepackage{url}
\usepackage{threeparttable}
\renewcommand{\Vec}[1]{\textrm{\boldmath $#1$}} 

\newcommand{\pt}[1]{\left(#1\right)} 

\def\pt#1{\left(#1\right)} 

\begin{document}

\title{Multi-Task Adversarial Training Algorithm for \\ Multi-Speaker Neural Text-to-Speech}

\author{%
\authorblockN{%
Yusuke Nakai, Yuki Saito, Kenta Udagawa, and Hiroshi Saruwatari
}
\authorblockA{%
Graduate School of Information Science and Technology, The University of Tokyo, Japan. \\
E-mail: nakai-yusuke@g.ecc.u-tokyo.ac.jp, yuuki\_saito@ipc.i.u-tokyo.ac.jp}
}

\maketitle
\thispagestyle{empty}

\begin{abstract}
We propose a novel training algorithm for a multi-speaker neural text-to-speech (TTS) model based on multi-task adversarial training. A conventional generative adversarial network (GAN)-based training algorithm significantly improves the quality of synthetic speech by reducing the statistical difference between natural and synthetic speech. However, the algorithm does not guarantee the generalization performance of the trained TTS model in synthesizing voices of unseen speakers who are not included in the training data. Our algorithm alternatively trains two deep neural networks: multi-task discriminator and multi-speaker neural TTS model (i.e., generator of GANs). The discriminator is trained not only to distinguish between natural and synthetic speech but also to verify the speaker of input speech is existent or non-existent (i.e., newly generated by interpolating seen speakers' embedding vectors). Meanwhile, the generator is trained to minimize the weighted sum of the speech reconstruction loss and adversarial loss for fooling the discriminator, which achieves high-quality multi-speaker TTS even if the target speaker is unseen. Experimental evaluation shows that our algorithm improves the quality of synthetic speech better than a conventional GANSpeech algorithm.
\end{abstract}

\section{Introduction}\label{sect:intro}

Multi-speaker text-to-speech (TTS)~\cite{hojo18spk_code} is a technology for synthesizing various speakers' voices from given text using a computer. This technology can extend the application range of TTS technology to personalized voice assistance and data augmentation for speech-discriminative tasks~\cite{ueno19icassp,huang21slt}. State-of-the-art multi-speaker neural TTS, i.e., a machine-learning-based framework for training deep neural networks (DNNs)~\cite{zen13dnn,oord16wavenet} that represent the TTS mapping, has enabled synthesizing voices of various speakers comparable to human speech~\cite{casanova22}. Such quality improvement has mainly arisen from developments of epoch-making DNN architectures (e.g., Transformer~\cite{vaswani17}), well-designed and sufficiently large multi-speaker speech corpora~\cite{zen19,takamichi20ast,shi21}, accurate speaker-identity modeling~\cite{jia18,cooper20zero_shot}, and sophisticated deep generative models~\cite{goodfellow14,kingma13vae,ho20ddpm}. Well-trained multi-speaker neural TTS models can be also applied to practical TTS scenarios such as voice cloning (i.e., reproducing an unseen speaker's voice with few utterances)~\cite{arik18}. This paper focuses on the development of an algorithm for training a high-fidelity multi-speaker neural TTS model.

Generative adversarial networks (GANs)~\cite{goodfellow14} are powerful deep generative models that have significantly improved the quality of synthetic speech in various TTS-related tasks, such as statistical parametric TTS~\cite{saito18taslp,kaneko17advps}, end-to-end TTS~\cite{donahue21,kim21vits}, and neural vocoding~\cite{kumar19,yamamoto20pwg}. The quality improvement derives from their frameworks to adversarially train two DNNs: discriminator and TTS model (i.e., generator), for learning the distribution of real data. Specifically, the discriminator is trained to distinguish between synthetic and natural speech and approximate the divergence (i.e., statistical difference) between them. Meanwhile, the generator is trained to minimize the approximated divergence by fooling the trained discriminator. This framework can be introduced into the TTS training by defining the objective function as the weighted sum of speech reconstruction loss (e.g., L1 or L2 loss between natural and generated speech parameters) and the adversarial loss for causing the discriminator to misclassify synthetic speech as natural. As a result, synthetic speech generated by the trained TTS model successfully reproduces the fine structures of natural speech parameters that tend to be over-smoothed by learning to minimize the speech reconstruction loss only~\cite{toda07_MLVC}.

One can extend the GAN-based training algorithm for multi-speaker neural TTS by conditioning the discriminator and TTS model by speaker representation. Such representation can be obtained as a one-hot speaker code~\cite{hojo18spk_code}, trainable lookup embedding~\cite{arik17}, or intermediate vector of a speaker encoder pretrained on speaker-discriminative tasks (e.g., d-vector~\cite{variani14dvecs} and x-vector~\cite{snyder18x_vector}). Zhao et al.~\cite{zhao18wgan-tts} proposed an algorithm to train a multi-speaker neural TTS model conditioned by the speaker code and defined the training objective as the speech reconstruction in both mel-spectrogram and waveform domains and the adversarial loss of the Wasserstein GAN~\cite{arjovsky17} with gradient penalty~\cite{gulrajani17}. Kanagawa et al.~\cite{kanagawa19ssw} modified the training objective of a discriminator so that it can classify the speaker identity as well as distinguish natural/synthetic speech. Yang et al.~\cite{yang21ganspeech} presented GANSpeech, state-of-the-art multi-speaker neural TTS model based on FastSpeech 2~\cite{ren21fs2} trained by a GAN-based algorithm incorporating the joint conditional and unconditional (JCU) discriminator~\cite{zhang18} and scaled feature matching loss~\cite{salimans16}. Although their algorithm achieved the quality of synthetic speech approaching that of natural speech, it does not guarantee the generalization performance of the trained TTS model in synthesizing voices of unseen speakers who are not included in the training data. The main reason is that the GAN-based algorithm regards the distribution of natural speech uttered by \textit{seen} speakers as the target to be trained and never considers whether the TTS model can synthesize realistic voices of \textit{unseen} speakers.

We propose a novel GAN-based multi-task training algorithm for a multi-speaker neural TTS model that can synthesize high-quality voices of unseen speakers. Like a GAN-based algorithm, our algorithm alternatively trains two DNNs: multi-task discriminator and multi-speaker neural TTS model. The discriminator aims to not only distinguish between natural and synthetic speech but also verify the speaker of input speech is existent or non-existent (i.e., newly generated by interpolating seen speakers' embedding vectors). Meanwhile, the TTS model tries to fool the discriminator by minimizing the weighted sum of the speech reconstruction loss and adversarial loss for fooling the discriminator. As a result, we can expect the trained TTS model to synthesize high-quality multi-speaker voices even if the target speaker is unseen. Experimental evaluation shows that our algorithm improves the quality of synthetic speech better than a conventional algorithm used in the training of GANSpeech~\cite{yang21ganspeech}.

\section{Baseline Methods}\label{sect:baseline}

\subsection{FastSpeech 2}\label{subsect:fs2}

FastSpeech 2~\cite{ren21fs2} is a non-autoregressive end-to-end TTS model widely used as the backbone of various TTS methods~\cite{yang21ganspeech,nishimura22interspeech}. It performs 1) phoneme duration prediction and sequence alignment using a duration predictor and a length regulator, 2) speech feature prediction using multiple variance adaptors, and 3) mel-spectrogram prediction using a mel-spectrogram decoder. A trained neural vocoder takes the predicted mel-spectrogram as input to synthesize a speech waveform. Although the phoneme alignment information is required in advance for the training, this TTS model can avoid some crucial inference errors (e.g., looping or skipping of text in synthetic speech) that often occur in autoregressive neural TTS methods such as Tacotron2~\cite{shen18}.

Let $\mathcal{X}$ and $\mathcal{Y}$ be a text set consisting of $N$ phoneme sequences $\{\Vec{x}_n\}_{n=1}^{N}$ and a speech set consisting of $N$ speech parameter sequences $\{\Vec{y}_n\}_{n=1}^{N}$, respectively. A speech parameter sequence contains phoneme durations $\Vec{y}_{\rm dur}$, intermediate features $\Vec{y}_{\rm feat}$ (e.g., $F_0$ and energy) predicted by the variance adaptors, and mel-spectrograms $\Vec{y}_{\rm mel}$. A FastSpeech 2-based TTS model $G(\Vec{\cdot})$ parameterized by $\Vec{\theta}^{\mathrm{(G)}}$ is trained to predict parameters of synthetic speech $\Vec{\hat y}_{*}$ from $\Vec{x}$. The objective function for training the model is given as:
\begin{align}
L_{\mathrm{FS2}}^{\mathrm{(G)}}\pt{\mathcal{X}, \mathcal{Y}}
& = \mathrm{MSE}\pt{\Vec{y}_{\rm dur}, \Vec{\hat y}_{\rm dur}} 
  + \mathrm{MSE}\pt{\Vec{y}_{\rm feat}, \Vec{\hat y}_{\rm feat}} \nonumber \\
& + \mathrm{MAE}\pt{\Vec{y}_{\rm mel}, \Vec{\hat y}_{\rm mel}}, \label{eq:L_FS2}
\end{align}
where $\mathrm{MAE}(\Vec{\cdot})$ and $\mathrm{MSE}(\Vec{\cdot})$ denote the mean absolute error and mean squared error between two given features, respectively. The model's parameters $\Vec{\theta}^{\mathrm{(G)}}$ are optimized by using the mini-batch stochastic gradient descent (SGD) $\Vec{\theta}^{\mathrm{(G)}} \leftarrow \Vec{\theta}^{\mathrm{(G)}} - \eta \Vec{\nabla}_{\Vec{\theta}^{\mathrm{(G)}}}L_{\mathrm{FS2}}^{\mathrm{(G)}}$ with a learning rate $\eta > 0$.

\subsection{Transfer-Learning-Based Multi-Speaker Neural TTS}\label{subsect:tl_mstts}

We adopt Jia et al.'s method~\cite{jia18} for building our baseline multi-speaker neural TTS model. In the training phase, a DNN-based speaker encoder that extracts speaker embedding from an input speech waveform is first trained on the objective function of speaker verification (e.g., minimizing the generalized end-to-end loss~\cite{wan18ge2e}). A multi-speaker neural TTS model is then trained to generate speech parameters from their corresponding phoneme sequence and speaker embedding $\Vec{z}$. In the inference phase, the embedding of a target speaker, who may be unseen during the training, is first extracted from reference speech uttered by the speaker. Then, the extracted speaker embedding and input text are fed into the trained TTS model to generate the target speaker's mel-spectrogram.

\subsection{GANSpeech}\label{subsect:ganspeech}

GANSpeech~\cite{yang21ganspeech} is the improved version of multi-speaker neural TTS based on FastSpeech 2, which incorporates the GAN-based training algorithm to achieve high-quality TTS. 

\subsubsection{JCU Discriminator}\label{subsubsect:jcu_disc}

A JCU discriminator $D(\Vec{\cdot})$ parameterized by $\Vec{\theta}^{\mathrm{(D)}}$ is designed to capture general characteristics of natural mel-spectrograms and speaker-specific features of them. Specifically, the discriminator consists of three sub-modules\footnote{We omit a fully-connected layer to transform speaker embedding for simplicity.}: 1) shared layers $D_{\mathrm{S}}(\Vec{\cdot})$, 2) conditional layers $D_{\mathrm{C}}(\Vec{\cdot})$, and 3) unconditional layers $D_{\mathrm{U}}(\Vec{\cdot})$, to output conditional and unconditional true/fake predictions, $t_{\mathrm{C}}$ and $t_{\mathrm{U}}$, from a mel-spectrogram $\Vec{y}_{\mathrm{mel}}$ and speaker embedding $\Vec{z}$. First, a hidden vector $\Vec{h}_{\mathrm{S}}$ is extracted by the shared layers from an input natural mel-spectrogram $\Vec{y}_{\mathrm{mel}}$ as $\Vec{h}_{\mathrm{S}} = D_{\mathrm{S}}(\Vec{y}_{\mathrm{mel}})$. Then, the conditional and unconditional predictions are obtained as $t_{\mathrm{C}} = D_{\mathrm{C}}(\Vec{h}_{\mathrm{S}}, \Vec{z})$ and $t_{\mathrm{U}} = D_{\mathrm{U}}(\Vec{h}_{\mathrm{S}})$, respectively. The same procedure is taken for a synthetic mel-spectrogram $\Vec{\hat y}_{\mathrm{mel}} = G(\Vec{x})$, and predictions $\hat{t}_{\mathrm{C}}$ and $\hat{t}_{\mathrm{U}}$ are obtained.

The discriminator is trained to distinguish natural speech $\Vec{y}_{\mathrm{mel}}$ from synthetic speech $\Vec{\hat y}_{\mathrm{mel}}$, with or without the conditioning by speaker embedding. The objective function is defined as follows:
\begin{align}\label{eq:GAN_D}
\begin{split}
L_{\rm GAN}^{\rm (D)}\pt{\mathcal{X}, \mathcal{Y}}
&= \frac{1}{2} \pt{ \pt{t_{\mathrm{C}} - 1}^2 + \pt{t_{\mathrm{U}} - 1}^2 } \\
&+ \frac{1}{2} \pt{ \hat{t}_{\mathrm{C}}^2 + \hat{t}_{\mathrm{U}}^2 }.
\end{split}
\end{align}

\subsubsection{Adversarial Loss}\label{subsubsect:adv_loss}

The TTS model is trained to make the discriminator misclassify synthetic speech as natural. The adversarial loss to cause the misclassification is defined as follows:
\begin{align}\label{eq:GAN_G}
L_{\rm GAN}^{\rm (G)}\pt{\mathcal{X}, \mathcal{Y}}
&= \frac{1}{2} \pt{ \pt{\hat{t}_{\mathrm{C}} - 1}^2 + \pt{\hat{t}_{\mathrm{U}} - 1}^2 }.
\end{align}

\begin{figure*}[tb]
\centering
\begin{minipage}[t]{0.7\linewidth}
\includegraphics[width=5.6in]{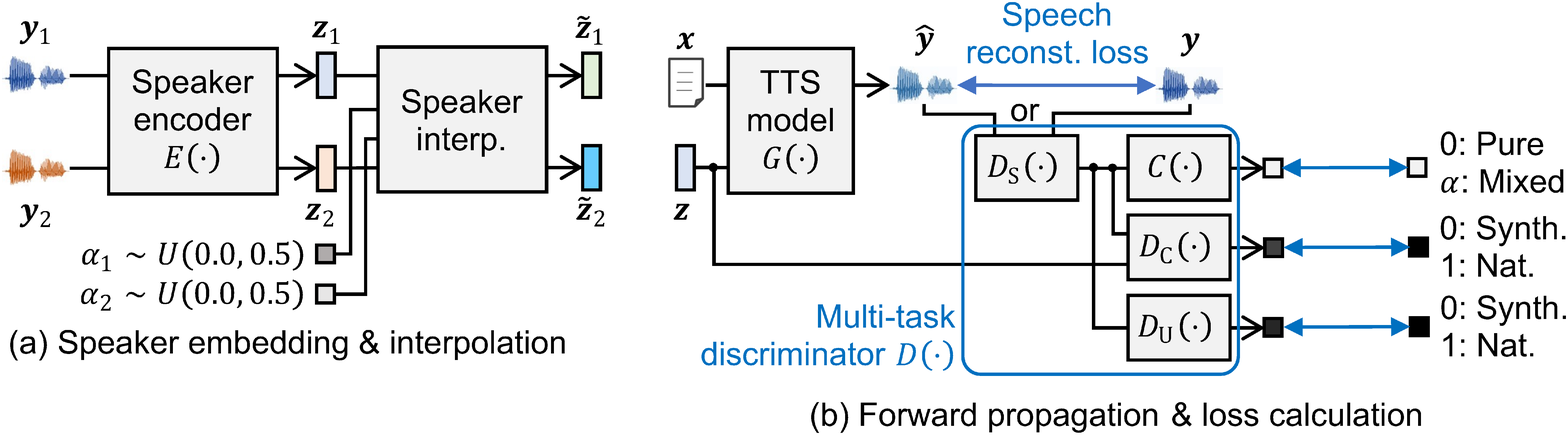}
\caption{Overview of proposed algorithm. The variables $\Vec{x}$, $\Vec{y}$, and $\Vec{z}$ denote text, speech, and speaker embedding, respectively. The interpolation coefficient $\alpha$ is sampled from the uniform distribution $U(0.0, 0.5)$.}\label{fig:prop}
\parbox{6.5cm}{\small \hspace{1.5cm} }
\end{minipage}
\end{figure*}

\subsubsection{Scaled Feature Matching Loss}\label{subsubsect:scld_fm_loss}

The feature matching loss~\cite{gulrajani17} is a well-known strategy to stabilize the GAN training and improve the quality of generated data. The core idea is to make the hidden vectors of the discriminator extracted by generated data $\Vec{\hat h}_{l}$ close to those of real data $\Vec{h}_{l}$ by minimizing the following loss:
\begin{align}\label{eq:GAN_FM}
L_{\rm FM}^{\rm (G)}\pt{\mathcal{X}, \mathcal{Y}}
&= \sum_{l=1}^{L} \frac{1}{N_{l}} \lVert \Vec{\hat h}_{l} - \Vec{h}_{l} \rVert_{1},
\end{align}
where $N_{l}$ denotes the number of elements in the $l$th hidden layer. In the GANSpeech training, a hyperparameter $\lambda_{\rm FM}$ to control the importance of feature matching is dynamically adjusted in accordance with the magnitude of reconstruction loss as $\lambda_{\rm FM} \leftarrow L_{\rm FS2}^{\rm (G)} / L_{\rm FM}^{\rm (G)}$.

\subsubsection{GANSpeech Algorithm}\label{subsubsect:alg_ganspeech}

The GANSpeech algorithm consists of three phases: 1) TTS model pretraining, 2) discriminator update, and 3) TTS model update. First, the TTS model is pretrained by minimizing the reconstruction loss $L_{\rm FS2}^{\rm (G)}$ only. Then, the two DNNs, $D(\Vec{\cdot})$ and $G(\Vec{\cdot})$, are alternatively trained. Specifically, the discriminator is updated as 
$\Vec{\theta}^{\mathrm{(D)}} \leftarrow \Vec{\theta}^{\mathrm{(D)}} - \eta \Vec{\nabla}_{\Vec{\theta}^{\mathrm{(D)}}}L_{\mathrm{GAN}}^{\mathrm{(D)}}$,
and the TTS model is trained to fool the updated discriminator as
$\Vec{\theta}^{\mathrm{(G)}} \leftarrow \Vec{\theta}^{\mathrm{(G)}} - \eta \Vec{\nabla}_{\Vec{\theta}^{\mathrm{(G)}}}L_{\mathrm{Total}}^{\mathrm{(G)}}$,
where the total loss for training the TTS model is given as $L_{\rm Total}^{\rm (G)} = L_{\rm FS2}^{\rm (G)} + L_{\rm GAN}^{\rm (G)} + \lambda_{\rm FM} L_{\rm FM}^{\rm (G)}$.

\section{Propose Algorithm}\label{sect:prop}

As mentioned in Section~\ref{sect:intro}, the GAN-based training algorithm for multi-speaker neural TTS does not guarantee that the trained model can synthesize high-quality speech of unseen speakers. To deal with the issue, we propose a novel algorithm for a multi-speaker neural TTS model to improve the quality of unseen speaker's synthetic speech.

\subsection{Adversarially Constrained Autoencoder Interpolation (ACAI)}\label{subsect:acai}
The ACAI algorithm~\cite{berthelot19} aims to learn an autoencoder that can generate realistic data from the convex combination of two latent variables. The encoder $E(\Vec{\cdot})$ extracts a latent variable (i.e., embedding) $\Vec{z}$ from input data $\Vec{y}$, and the decoder $G(\Vec{\cdot})$ reconstructs the data $\Vec{y}$ from $\Vec{e}$. In addition to this autoencoding process, $\Vec{\hat y} = G(E(\Vec{y}))$, a critic $C(\Vec{\cdot})$ regularizes the autoencoder so that interpolation between two embedding vectors also provides realistic data by decoding. This regularization is formulated as the adversarial training between the autoencoder and critic. The critic is trained to minimize the following loss function:
\begin{align}\label{eq:ACAI_C}
L_{\rm ACAI}^{\rm (C)}\pt{\mathcal{Y}}
&= \lVert C\pt{\Vec{y}} \rVert_2^2 + \lVert C\pt{\Vec{\tilde y}_{\alpha}} - \alpha \rVert_2^2,
\end{align}
where $\Vec{\tilde y}$ is decoded from the interpolated embedding with an interpolation coefficient $\alpha \sim U(0.0, 0.5)$ as $G(\alpha E(\Vec{y}_1) + (1 - \alpha) E(\Vec{y}_2))$. The first and second terms enforce that the critic outputs $\alpha = 0$ for pure data without interpolation\footnote{To stabilize the training in initial stage, the original algorithm mixes true data and reconstructed data with a coefficient $\gamma$ and regards it as the non-interpolated data~\cite{berthelot19}. Our algorithm omits this process because it first pretrains the decoder (i.e., TTS model) without any adversarial constraints.} and predicts the mixing coefficient of interpolated data, respectively. Meanwhile, the autoencoder is trained to minimize the following loss function:
\begin{align}\label{eq:ACAI_EG}
L_{\rm ACAI}^{\rm (E,G)}\pt{\mathcal{Y}}
&= \lVert \Vec{y} - G\pt{E\pt{\Vec{y}}} \rVert_2^2 + \lambda_{\rm ACAI} \lVert C\pt{\Vec{\tilde y}} \rVert_2^2,
\end{align}
where $\lambda_{\rm ACAI}$ is a hyperparameter to control the effect of the second term that makes the critic recognize the interpolated data as non-interpolated one.

\subsection{Multi-Task Adversarial Training Algorithm}

We introduce the ACAI-derived regularization term to the GANSpeech algorithm for training a multi-speaker neural TTS model. Specifically, we regard a pre-trained speaker encoder and the TTS model as an encoder $E(\Vec{\cdot})$ and decoder $G(\Vec{\cdot})$ used in the ACAI algorithm, respectively. Furthermore, we extend the JCU discriminator used in GANSpeech to a multi-task discriminator $D(\Vec{\cdot})$ that simultaneously predicts 1) true or fake of input mel-spectrogram with/without conditioning by speaker embedding and 2) coefficient of speaker interpolation. Figure~\ref{fig:prop} shows the overview of our algorithm. Our algorithm requires additional computation in training due to the multi-task discrimination by $D(\Vec{\cdot})$ but does not increase the inference time.

\subsubsection{Multi-Task Discriminator}\label{subsubsect:mtd}

Our multi-task discriminator consists of four sub-modules: the three layers included in the JCU discriminator of GANSpeech (i.e., $D_{\mathrm{S}}(\Vec{\cdot})$, $D_{\mathrm{C}}(\Vec{\cdot})$, and $D_{\mathrm{U}}(\Vec{\cdot})$) and ACAI-derived critic layers $C(\Vec{\cdot})$. The critic layers predicts the speaker interpolation coefficient of input mel-spectrogram $\hat{\alpha}$ from the hidden vector $\Vec{h}^{\rm (S)}$ extracted by the shared layers $D_{\mathrm{S}}(\Vec{\cdot})$.

\subsubsection{Forward Propagation}\label{subsubsect:prop_fwd}

Let $\Vec{X} = [ \Vec{x}_1, \Vec{x}_2, \ldots, \Vec{x}_{M} ] \in \mathcal{X}$ and $\Vec{Y} = [ \Vec{y}_1, \Vec{y}_2, \ldots, \Vec{y}_{M} ] \in \mathcal{Y}$ be mini-batches containing $M$ phoneme sequences and speech parameters, respectively. In the forward propagation during the training, our algorithm takes the following processes.

{\bf Speaker embedding extraction and interpolation:} The speaker encoder first extracts speaker embeddings $\Vec{Z} = [ \Vec{z}_1, \Vec{z}_2, \ldots, \Vec{z}_M ]$ from the speech parameters as $\Vec{Z} = E(\Vec{Y})$. Then, interpolation coefficients for each embedding $\Vec{\alpha} = [ \alpha_1, \alpha_2, \ldots, \alpha_M ]^{\top}$ are sampled from $\mathrm{Uniform}(0.0, 0.5)$. Finally, interpolated speaker embedding $\Vec{\tilde Z} = [ \Vec{\tilde z}_1, \Vec{\tilde z}_2, \ldots, \Vec{\tilde z}_M ]$ are calculated by using the convex combinations of the original embeddings $\Vec{Z}$ and their reversed vectors, i.e.,
$\Vec{\tilde z}_{m} = \alpha_{m} \Vec{z}_{m} + (1 - \alpha_{m}) \Vec{z}_{M-m+1}$ for $m = 1, 2, \ldots, M$.

{\bf Multi-speaker TTS:} The TTS model generates $2M$ speech parameters, $\Vec{\hat Y} = [ \Vec{\hat y}_1, \Vec{\hat y}_2, \ldots, \Vec{\hat y}_{M} ]$ and $\Vec{\tilde Y} = [ \Vec{\tilde y}_1, \Vec{\tilde y}_2, \ldots, \Vec{\tilde y}_{M} ]$, from input texts $\Vec{X}$ paired with original speaker embeddings $\Vec{Z}$ and interpolated ones $\Vec{\tilde Z}$ as $\Vec{\hat Y} = G(\Vec{X}, \Vec{Z})$ and $\Vec{\tilde Y} = G(\Vec{X}, \Vec{\tilde Z})$, respectively. The natural and generated speech parameters are used for the computation of speech reconstruction loss $L_{\rm FS2}^{\rm (G)}$ shown in Eq.~(\ref{eq:L_FS2}).

{\bf Multi-task discrimination:} The discriminator outputs six scalar values: true/fake predictions for \{ natural, generated \} mel-spectrograms \{ with, without \} conditioning by speaker embeddings and speaker interpolation coefficients for \{ pure, speaker-interpolated \} mel-spectrograms. These values are used for the computation of the objective function $L_{\rm MT}^{\rm (D)} = L_{\rm GAN}^{\rm (D)} + L_{\rm ACAI}^{\rm (D)}$, i.e., the sum of Eqs.~(\ref{eq:GAN_D}) and (\ref{eq:ACAI_C}).

\subsubsection{Backward Propagation and Model Update}\label{subsubsect:prop_bkwd}

The discriminator is first updated to minimize the multi-task discrimination loss $L_{\rm MT}^{\rm (D)}$ as $\Vec{\theta}^{\mathrm{(D)}} \leftarrow \Vec{\theta}^{\mathrm{(D)}} - \eta \Vec{\nabla}_{\Vec{\theta}^{\mathrm{(D)}}}L_{\mathrm{MT}}^{\mathrm{(D)}}$. The TTS model is then updated to fool the updated discriminator by minimizing $L_{\rm MT}^{\rm (G)} = L_{\rm Total}^{\rm (G)} + \lambda_{\rm ACAI} \lVert C(D_{\rm S}(\Vec{\tilde y})) \rVert_2^2$ as $\Vec{\theta}^{\mathrm{(G)}} \leftarrow \Vec{\theta}^{\mathrm{(G)}} - \eta \Vec{\nabla}_{\Vec{\theta}^{\mathrm{(G)}}}L_{\mathrm{MT}}^{\mathrm{(G)}}$. Note that the speaker encoder is not updated because we focus on the effect of adversarial training only, not the fine-tuning of the encoder.

\section{Experiments}\label{sect:exp}
\subsection{Experimental Conditions}\label{subsect:exp_cond}

For training a speaker encoder, we used the Corpus of Spontaneous Japanese (CSJ)~\cite{maekawa03csj} containing 660 hours of speech data from 1,417 Japanese speakers (947 men and 470 women). The CSJ speech data were resampled to 16~kHz, and the frame shift was set to 10~ms. For training, validating, and evaluating a multi-speaker neural TTS model, we used the ``parallel100'' subset of the Japanese Versatile Speech (JVS) corpus~\cite{takamichi20ast}. The subset contains 22 hours of speech data from 100 Japanese speakers (49 men and 51 women; 100 sentences per speaker). The ratio for each of training, validation, and test data was 0.8, 0.1, and 0.1, respectively. Referring to Udagawa et al.'s study~\cite{udagawa22interspeech}, we regarded the following four speakers: ``jvs078,'' ``jvs060,'' ''jvs005,'' and ``jvs010'' as unseen speakers during the training. The JVS speech data were resampled to 22.05~kHz to match the settings of neural vocoding, and the frame shift was set to 12~ms.

We used the open source implementation of FastSpeech 2~\cite{ren21fs2} published by Wataru-Nakata\footnote{\url{https://github.com/Wataru-Nakata/FastSpeech2-JSUT}} for building our TTS model. The TTS model predicted 80-dimensional mel-spectrogram from Japanese phonemes with the aid of the variance adaptors that predicted the $F_0$ and energy of synthetic speech. The $F_0$ was estimated using the WORLD vocoder~\cite{morise16world,morise16d4c}. In the FastSpeech 2 training, we used the Adam optimizer~\cite{kingma14adam} with 4,000 Warmup~\cite{goyal17accurate} steps, initial learning rate of 0.0625, batch size of 8, and 30,000 training steps.

The DNN architecture architecture of a speaker encoder was the same as that used in Jia et al.'s method~\cite{jia18}. The dimensionality of speaker embeddings was 256. In the speaker-encoder training, we used the Adam optimizer with a learning rate of 0.0001, batch size of 8, and 1,000,000 training step.

Our multi-task discriminator had a structure of its DNN architecture similar to that of the JCU discriminator in GANSpeech~\cite{yang21ganspeech}. Specifically, it consists a stack of 1D convolutional (Conv1D) layers with the parameters of convolution operations $\mathrm{Conv1D}(c, k, s)$, where $c$, $k$, and $s$ denote the number of output channels, kernel size, and stride width, respectively. The activation functions for hidden and output layers were leaky ReLU~\cite{maas13} and sigmoid, respectively. The $D_{\rm S}(\Vec{\cdot})$ had three Conv1D layers: $\mathrm{Conv1D}(64, 3, 1)$, $\mathrm{Conv1D}(128, 5, 2)$, and $\mathrm{Conv1D}(512, 5, 2)$. The other sub-modules, $D_{\rm C}(\Vec{\cdot})$, $D_{\rm U}(\Vec{\cdot})$, and $C(\Vec{\cdot})$, were all consisted of two Conv1D layers:  $\mathrm{Conv1D}(128, 5, 2)$ and $\mathrm{Conv1D}(1, 3, 1)$, except for conditioning by speaker embedding in $D_{\rm C}(\Vec{\cdot})$.

The neural vocoder for synthesizing speech waveform was the ``generator\_universal model'' of HiFi-GAN~\cite{kong20} included in the FastSpeech 2 repository published by ming024\footnote{\url{https://github.com/ming024/FastSpeech2}}. We did not finetune the HiFi-GAN on Japanese speech data.

\subsection{Subjective Evaluations}\label{subsect:sbj_eval}

We evaluated the performance of our algorithm on two TTS tasks: 1) TTS for seen speakers and 2) voice cloning for unseen speakers. We compared the following three algorithms:
\begin{enumerate}
 \item 
 {\bf FS2}: Minimizing $L_{\rm FS2}^{\rm (G)}$ (Eq.~(\ref{eq:L_FS2})) only~\cite{ren21fs2}
 \item 
 {\bf GAN}: Minimizing $L_{\rm Total}^{\rm (G)} = L_{\rm FS2}^{\rm (G)} + L_{\rm GAN}^{\rm (G)} + \lambda_{\rm FM} L_{\rm FM}^{\rm (G)}$ (i.e., the GANSpeech algorithm~\cite{yang21ganspeech})
 \item 
 {\bf MT}: Minimizing $L_{\rm MT}^{\rm (G)} = L_{\rm Total}^{\rm (G)} + \lambda_{\rm ACAI} \lVert C(D_{\rm S}(\Vec{\tilde y})) \rVert_2^2$ (i.e., the proposed algorithm)
\end{enumerate}
We chose the hyperparameter $\lambda_{\rm ACAI} = 1.0$ empirically. Speech samples used for the evaluation are available on \url{http://sython.org/demo/nakai22apsipa/demo.html}.

\subsubsection{TTS for Seen Speakers}\label{subsubsect:seen_TTS}

We conducted the five-level mean opinion score (MOS) and degradation MOS (DMOS) tests to  evaluate the naturalness and speaker similarity of synthetic speech using the test data of 96 seen speakers, respectively. In the DMOS test, natural speech uttered by a speaker to be synthesized was used as the reference for evaluating the similarity. For each test, we recruited 50 listeners who evaluated the quality of speech samples using our crowdsourced subjective evaluation system. Each listener evaluated 15 speech samples whose speakers were randomized.

\begin{table}[tb]
\centering
\caption{Subjective evaluation results with their 95\% confidence intervals (TTS for seen speakers)}
\label{table:seen_tts}
\vspace{-5pt}
\begin{tabular}{c|c|c}
\hline 
\hline
Algorithm & Naturalness MOS & Speaker similarity DMOS \\
\hline
FS2 & 3.18 $\pm$ 0.12 & 3.57 $\pm$ 0.12 \\
GAN & 3.52 $\pm$ 0.12 & 3.79 $\pm$ 0.12 \\
MT &  \textbf{3.55 $\pm$ 0.12} & \textbf{3.87 $\pm$ 0.12} \\
\hline
\hline
\end{tabular}
\end{table}

Table~\ref{table:seen_tts} shows the evaluation results. From this table, we observe that 1) ``GAN'' and ``MT'' significantly outperform ``FS2'' and 2) ``MT'' achieves similar performance to ``GAN.''  These results demonstrate that our multi-task adversarial training algorithm can keep the performance of the conventional GANSpeech algorithm for TTS of seen speakers.

\subsubsection{Voice Cloning for Unseen Speakers}\label{subsubsect:unseen_TTS}

We also evaluated the performance of our algorithm for voice cloning of unseen speakers and conducted the MOS and DMOS tests. The conditions of the evaluation, i.e., the numbers of listeners and speech samples, were the same as those in Section~\ref{subsubsect:seen_TTS}. The speaker embeddings of the unseen speakers were estimated by utterance-wise.

\begin{table}[tb]
\centering
\caption{Subjective evaluation results with their 95\% confidence intervals (TTS for unseen speakers)}
\label{table:unseen_tts}
\vspace{-5pt}
\begin{tabular}{c|c|c}
\hline 
\hline
Algorithm & Naturalness MOS & Speaker similarity DMOS \\
\hline
FS2 & 3.13 $\pm$ 0.12 & 2.38 $\pm$ 0.12 \\
GAN & 3.38 $\pm$ 0.12 & 2.40 $\pm$ 0.13 \\
MT &  \textbf{3.50 $\pm$ 0.12} & \textbf{2.48 $\pm$ 0.12} \\
\hline
\hline
\end{tabular}
\end{table}

Table~\ref{table:unseen_tts} shows the evaluation results. From this table, we confirm that ``MT'' achieves the highest MOS and DMOS values among the three algorithms, which suggests that our multi-task adversarial training algorithm can mitigate the degradation of speech naturalness in the voice cloning task and successfully improve the naturalness better than the conventional GANSpeech algorithm. However, we observe that there is a still large gap between the DMOS values of synthetic speech in seen speakers' TTS (the third column of Table~\ref{table:seen_tts}) and those of cloned speech. This result indicates that the ACAI-derived regularization is insufficient to diversify the variation of speakers and another approach such as speaker generation~\cite{stanton22} may be necessary for the absolute improvement of speaker similarity.

\section{Conclusion}\label{sect:concl}

We proposed a GAN-based multi-task training algorithm for a multi-speaker neural TTS model that can synthesize high-quality voices of unseen speakers. Our algorithm alternatively trains two DNNs: multi-task discriminator and multi-speaker neural TTS model. The discriminator aims to not only distinguish between natural and synthetic speech but also verify the speaker of input speech is existent or non-existent (i.e., newly generated by interpolating seen speakers' embedding vectors). Meanwhile, the TTS model tries to fool the discriminator by minimizing the weighted sum of the speech reconstruction loss and adversarial loss for fooling the discriminator. Experimental evaluation showed that our algorithm improved the quality of synthetic speech better than a conventional algorithm used in the training of GANSpeech. In future work, we will investigate how to improve the speaker similarity of synthetic speech in voice cloning and conduct TTS experiments for other languages.

\section*{Acknowledgment}
 
This work was supported by JST, Moonshot R\&D Grant Number JPMJPS2011 (algorithm development) and JSPS KAKENHI Grant Number 21K21305 (practical experiment).

\bibliographystyle{IEEEbib}
\bibliography{tts}

\begin{thebibliography}{10}

\bibitem{hojo18spk_code}
N.~Hojo, Y.~Ijima, and H.~Mizuno,
\newblock ``{DNN}-based speech synthesis using speaker codes,''
\newblock {\em IEICE Transactions on Information and Systems}, vol. E101-D, no.
  2, pp. 462--472, Feb. 2018.

\bibitem{ueno19icassp}
S.~Ueno, M.~Mimura, S.~Sakai, and T.~Kawahara,
\newblock ``Multi-speaker sequence-to-sequence speech synthesis for data
  augmentation in acoustic-to-word speech recognition,''
\newblock in {\em Proc. ICASSP}, Brighton, U. K., May 2019, pp. 6161--6165.

\bibitem{huang21slt}
J.~Pelecanos Y.~Huang, Y.~Chen and Q.~Wang,
\newblock ``Synth2{A}ug: {C}ross-domain speaker recognition with {TTS}
  synthesized speech,''
\newblock in {\em Proc. SLT}, Shenzhen, China, Jan. 2021, pp. 316--322.

\bibitem{zen13dnn}
H.~Zen, A.~Senior, and M.~Schuster,
\newblock ``Statistical parametric speech synthesis using deep neural
  networks,''
\newblock in {\em Proc. ICASSP}, Vancouver, Canada, May 2013, pp. 7962--7966.

\bibitem{oord16wavenet}
A.~Oord, S.~Dieleman, H.~Zen, K.~Simonyan, O.~Vinyals, A.~Graves,
  N.~Kalchbrenner, A.~Senior, and K.~Kavukcuoglu,
\newblock ``Wave{N}et: {A} generative model for raw audio,''
\newblock {\em arXiv}, vol. abs/1609.03499, 2016.

\bibitem{casanova22}
E.~Casanova, J.~Weber, C.~D. Shulby, A.~C. Junior, E.~G{\"o}lge, and M.~A.
  Ponti,
\newblock ``{Y}our{TTS}: Towards zero-shot multi-speaker {TTS} and zero-shot
  voice conversion for everyone,''
\newblock in {\em Proc. ICML}, Baltimore, U.S.A., Jul. 2022, pp. 2709--2720.

\bibitem{vaswani17}
N.~Parmar J. Uszkoreit L. Jones A. N. Gomez L.~Kaiser A.~Vaswani, N.~Shazeer
  and I.~Polosukhin,
\newblock ``Attention is all you need,''
\newblock in {\em Proc. NIPS}, Long Beach, U.S.A., Dec. 2017.

\bibitem{zen19}
H.~Zen, V.~Dang, R.~Clark, Y.~Zhang, R.~J. Weiss, Y.~Jia, Z.~Chen, and Y.~Wu,
\newblock ``Libri{TTS}: A corpus derived from {L}ibri{S}peech for
  text-to-speech,''
\newblock in {\em Proc. INTERSPEECH}, Graz, Austria, Sep. 2019, pp. 1526--1530.

\bibitem{takamichi20ast}
S.~Takamichi, R.~Sonobe, K.~Mitsui, Y.~Saito, T.~Koriyama, N.~Tanji, and
  H.~Saruwatari,
\newblock ``{JSUT and JVS}: Free {J}apanese voice corpora for accelerating
  speech synthesis research,''
\newblock {\em Acoustical Science and Technology}, vol. 41, no. 5, pp.
  761--768, Sep. 2020.

\bibitem{shi21}
Y.~Shi, H.~Bu, X.~Xu, S.~Zhang, and M.~Li,
\newblock ``{AISHELL-3}: A multi-speaker {M}andarin {TTS} corpus,''
\newblock in {\em Proc. INTERSPEECH}, Brno, Czech Republic, Sep. 2021, pp.
  2756--2760.

\bibitem{jia18}
Y.~Jia, Y.~Zhang, R.~Weiss, Q.~Wang, J.~Shen, F.~Ren, Z.~Chen, P.~Nguyen,
  R.~Pang, I.~L. Moreno, and Y.~Wu,
\newblock ``Transfer learning from speaker verification to multispeaker
  text-to-speech synthesis,''
\newblock in {\em Proc. NeurIPS}, Montreal, Canada, Dec. 2018, pp. 4480--4490.

\bibitem{cooper20zero_shot}
E.~Cooper, C.-I Lai, Y.~Yasuda, F.~Fang, X.~Wang, N.~Chen, and J.~Yamagishi,
\newblock ``Zero-shot multi-speaker text-to-speech with state-of-the-art neural
  speaker embeddings,''
\newblock in {\em Proc. ICASSP}, Barcelona, Spain, May 2020, pp. 6184--6188.

\bibitem{goodfellow14}
I.~Goodfellow, J.~Pouget-Abadie, M.~Mirza, B.~Xu, D.~Warde-Farley, S.~Ozair,
  A.~Courville, and Y.~Bengio,
\newblock ``Generative adversarial nets,''
\newblock in {\em Proc. NIPS}, Montreal, Canada, Dec. 2014, pp. 2672--2680.

\bibitem{kingma13vae}
D.~P. Kingma and M.~Welling,
\newblock ``Auto-encoding variational {B}ayes,''
\newblock {\em arXiv}, vol. abs/1312.6114, 2013.

\bibitem{ho20ddpm}
J.~Ho, A.~Jain, and P.~Abbeel,
\newblock ``Denoising diffusion probabilistic models,''
\newblock in {\em Proc. NeurIPS}, Vancouver, Canada, Dec. 2020.

\bibitem{arik18}
S.~O. Arik, J.~Chen, K.~Peng, W.~Ping, and Y.~Zhou,
\newblock ``Neural voice cloning with a few samples,''
\newblock in {\em Proc. NeurIPS}, Montreal, Canada, Dec. 2018, pp.
  10019--10029.

\bibitem{saito18taslp}
Y.~Saito, S.~Takamichi, and H.~Saruwatari,
\newblock ``Statistical parametric speech synthesis incorporating generative
  adversarial networks,''
\newblock {\em IEEE/ACM Transactions on Audio, Speech, and Language
  Processing}, vol. 26, no. 1, pp. 84--96, Jan. 2018.

\bibitem{kaneko17advps}
T.~Kaneko, H.~Kameoka, N.~Hojo, Y.~Ijima, K.~Hiramatsu, and K.~Kashino,
\newblock ``Generative adversarial network-based postfilter for statistical
  parametric speech synthesis,''
\newblock in {\em Proc. ICASSP}, New Orleans, U.S.A., Mar. 2017, pp.
  4910--4914.

\bibitem{donahue21}
J.~Donahue, S.~Dieleman, M.~Binkowski, E.~Elsen, and K.~Simonyan,
\newblock ``End-to-end adversarial text-to-speech,''
\newblock in {\em Proc. ICLR}, Vienna, Austria, May 2021.

\bibitem{kim21vits}
J.~Kim, J.~Kong, and J.~Son,
\newblock ``Conditional variational autoencoder with adversarial learning for
  end-to-end text-to-speech,''
\newblock in {\em Proc. ICML}, Virtual Conference, Jun. 2021, pp. 5530--5540.

\bibitem{kumar19}
K.~Kumar, R.~Kumar, T.~de~Boissiere, L.~Gestin, W.~Z. Teoh, J.~Sotelo,
  A.~de~Brébisson, Y.~Bengio, and A.~C. Courville,
\newblock ``{MelGAN}: {G}enerative adversarial networks for conditional
  waveform synthesis,''
\newblock in {\em Proc. NeurIPS}, Vancouver, Canada, Dec. 2019, pp.
  14881--14892.

\bibitem{yamamoto20pwg}
R.~Yamamoto, E.~Song, and J.~Kim,
\newblock ``Parallel {W}ave{GAN}: a fast waveform generation model based on
  generative adversarial networks with multi-resolution spectrogram,''
\newblock in {\em Proc. ICASSP}, Barcelona, Spain, May 2020, pp. 6199--6203.

\bibitem{toda07_MLVC}
T.~Toda, A.~W. Black, and K.~Tokuda,
\newblock ``Voice conversion based on maximum likelihood estimation of spectral
  parameter trajectory,''
\newblock {\em IEEE Transactions on Audio, Speech, and Language Processing},
  vol. 15, no. 8, pp. 2222--2235, Nov. 2007.

\bibitem{arik17}
S.~O. Arik, G.~Diamosy, A.~Gibiansky, J.~Miller, K.~Peng, W.~Ping, J.~Raiman,
  and Y.~Zhou,
\newblock ``Deep {V}oice 2: {M}ulti-speaker neural text-to-speech,''
\newblock in {\em Proc. NIPS}, Red Hook, New York, U.S.A., May 2017, pp.
  2966--2974.

\bibitem{variani14dvecs}
E.~Variani, X.~Lei, E.~McDermott, I.~L. Moreno, and J.~Gonzalez-Dominguez,
\newblock ``Deep neural networks for small footprint text-dependent speaker
  verification,''
\newblock in {\em Proc. ICASSP}, Florence, Italy, May 2014, pp. 4080--4084.

\bibitem{snyder18x_vector}
D.~Snyder, D.~Garcia-Romero, G.~Sell, D.~Povey, and S.~Khudanpur,
\newblock ``X-vectors: robust {DNN} embeddings for speaker recognition,''
\newblock in {\em Proc. ICASSP}, Alberta, Canada, Apr. 2018, pp. 5329--5333.

\bibitem{zhao18wgan-tts}
Y.~Zhao, S.~Takaki, H.-T. Luong, J.~Yamagishi, D.~Saito, and N.~Minematsu,
\newblock ``Wasserstein {GAN} and waveform loss-based acoustic model training
  for multi-speaker text-to-speech synthesis systems using a {W}ave{N}et
  vocoder,''
\newblock {\em IEEE Access}, vol. 6, pp. 60478--60488, Sep. 2018.

\bibitem{arjovsky17}
M.~Arjovsky, S.~Chintala, and L.~Bottou,
\newblock ``Wasserstein {GAN},''
\newblock {\em arXiv}, vol. abs/1701.07875, 2017.

\bibitem{gulrajani17}
I.~Gulrajani, F.~Ahmed, M.~Arjovsky, V.~Dumoulin, and A.~Courville,
\newblock ``Improved training of {W}asserstein {GAN}s,''
\newblock in {\em Proc. NIPS}, Long Beach, U.S.A., Dec. 2017.

\bibitem{kanagawa19ssw}
H.~Kanagawa and Y.~Ijima,
\newblock ``Multi-speaker modeling for {DNN}-based speech synthesis
  incorporating generative adversarial networks,''
\newblock in {\em Proc. SSW}, Vienna, Austria, Sep. 2019, pp. 40--44.

\bibitem{yang21ganspeech}
J.~Yang, J.-S. Bae, T.~Bak, Y.~Kim, and H.-Y. Cho,
\newblock ``{GANS}peech: {A}dversarial training for high-fidelity multi-speaker
  speech synthesis,''
\newblock in {\em Proc. INTERSPEECH}, Brno, Czech Republic, Sep. 2021, pp.
  2202--2206.

\bibitem{ren21fs2}
Y.~Ren, C.~Hu, X.~Tan, T.~Qin, S.~Zhao, Z.~Zhao, and T.-Y. Liu,
\newblock ``{FastSpeech} 2: {F}ast and high-quality end-to-end text to
  speech,''
\newblock in {\em Proc. ICLR}, Vienna, Austria, May 2021.

\bibitem{zhang18}
H.~Zhang, H.~Li T.~Xu, S, Zhang, X.~Wang, X.~Huang, and D.~N. Metaxas,
\newblock ``Stack{GAN}++: {R}ealistic image synthesis with stacked generative
  adversarial networks,''
\newblock {\em IEEE Transactions on Pattern Analysis and Machine Intelligence},
  vol. 41, no. 8, pp. 1947--1962, Jul. 2018.

\bibitem{salimans16}
T.~Salimans, I.~Goodfellow, W.~Zaremba, V.~Cheung, A.~Radford, and X.~Chen,
\newblock ``Improved techniques for training {GAN}s,''
\newblock in {\em Proc. NIPS}, Barcelona, Spain, Dec. 2016, pp. 2234--2242.

\bibitem{nishimura22interspeech}
Y.~Nishimura, Y.~Saito, S.~Takamichi, K.~Tachibana, and H.~Saruwatari,
\newblock ``Acoustic modeling for end-to-end empathetic dialogue speech
  synthesis using linguistic and prosodic contexts of dialogue history,''
\newblock in {\em Proc. INTERSPEECH}, Incheon, South Korea, Sep. 2022, pp.
  xxxx--xxxx,
\newblock (ACCEPTED).

\bibitem{shen18}
J.~Shen, R.~Pang, R.~J. Weiss, M.~Schuster, N.~Jaitly, Z.~Yang, Z.~Chen,
  Y.~Zhang, Y.~Wang, RJ~Skerry-Ryan, R.~A. Saurous, Y.~Agiomyrgiannakis, and
  Y.~Wu,
\newblock ``Natural {TTS} synthesis by conditioning {W}ave{N}et on mel
  spectrogram predictions,''
\newblock in {\em Proc. ICASSP}, Calgary, Canada, Apr. 2018, pp. 4779--4783.

\bibitem{wan18ge2e}
L.~Wan, Q.~Wang, A.~Papir, and I.~L. Moreno,
\newblock ``Generalized end-to-end loss for speaker verification,''
\newblock in {\em Proc. ICASSP}, Alberta, Canada, Apr. 2018, pp. 4879--4883.

\bibitem{berthelot19}
D.~Berthelot, C.~Raffel, A.~Roy, and I.~Goodfellow,
\newblock ``Understanding and improving interpolation in autoencoders via an
  adversarial regularizer,''
\newblock in {\em Proc. ICLR}, New Orleans, U.S.A., May 2019.

\bibitem{maekawa03csj}
K.~Maekawa,
\newblock ``Corpus of spontaneous {J}apanese: Its design and evaluation,''
\newblock in {\em Proc. SSPR}, Tokyo, Japan, Apr. 2003, pp. 7--12.

\bibitem{udagawa22interspeech}
K.~Udagawa, Y.~Saito, and H.~Saruwatari,
\newblock ``Human-in-the-loop speaker adaptation for {DNN}-based multi-speaker
  {TTS},''
\newblock in {\em Proc. INTERSPEECH}, Incheon, South Korea, Sep. 2022, pp.
  xxxx--xxxx,
\newblock (ACCEPTED).

\bibitem{morise16world}
M.~Morise, F.~Yokomori, and K.~Ozawa,
\newblock ``{WORLD}: a vocoder-based high-quality speech synthesis system for
  real-time applications,''
\newblock {\em IEICE Transactions on Information and Systems}, vol. E99-D, no.
  7, pp. 1877--1884, 2016.

\bibitem{morise16d4c}
M.~Morise,
\newblock ``{D4C}, a band-aperiodicity estimator for high-quality speech
  synthesis,''
\newblock {\em Speech Communication}, vol. 84, pp. 57--65, Nov. 2016.

\bibitem{kingma14adam}
D.~Kingma and B.~Jimmy,
\newblock ``Adam: A method for stochastic optimization,''
\newblock in {\em ar{X}iv preprint ar{X}iv:1412.6980}, 2014.

\bibitem{goyal17accurate}
Priya Goyal, Piotr Doll{\'a}r, Ross~B. Girshick, P.~Noordhuis, Lukasz
  Wesolowski, Aapo Kyrola, Andrew Tulloch, Y.~Jia, and Kaiming He,
\newblock ``Accurate, large minibatch {SGD}: Training imagenet in 1 hour,''
\newblock {\em arXiv}, vol. abs/1706.02677, 2017.

\bibitem{maas13}
A.~L. Maas, A.~Y. Hannun, and A.~Y. Ng,
\newblock ``Rectifier nonlinearities improve neural network acoustic models,''
\newblock in {\em Proc. ICML}, Atlanta, U.S.A., Jun. 2013.

\bibitem{kong20}
J.~Kong, J.~Kim, and J.~Bae,
\newblock ``{HiFi-GAN}: Generative adversarial networks for efficient and high
  fidelity speech synthesis,''
\newblock in {\em Proc. NeurIPS}, Vancouver, Canada, Dec. 2020.

\bibitem{stanton22}
D.~Stanton, M.~Shannon, S.~Mariooryad, RJ~Skerry-Ryan, E.~Battenberg, T.~Bagby,
  and D.~Kao,
\newblock ``Speaker generation,''
\newblock in {\em Proc. ICASSP}, Singapore, May 2022, pp. 7897--7901.

\end{thebibliography}
\end{document}